\documentclass[twocolumn,showpacs,preprintnumbers,amsmath,amssymb]{revtex4}

\usepackage{amsmath,amssymb,graphics,epsfig,subfigure,color,hyperref}
\usepackage{color}

\newcommand {\nn}    {\nonumber}

\begin{document}

\title{Tensor perturbations of $f(T)$-branes}

\author{Wen-Di Guo}
\author{Qi-Ming Fu}
\author{Yu-Peng Zhang}
\author{Yu-Xiao Liu\footnote{liuyx@lzu.edu.cn, corresponding author}}

\address{Institute of Theoretical Physics, Lanzhou University,\\
           Lanzhou 730000, People's Republic of China}

\begin{abstract}
We explore the tensor perturbation of the $f(T)$ brane embedded in an AdS$_5$ spacetime. With the transverse-traceless condition, we get the tensor perturbation equation of the $f(T)$ brane and show that  the stability of this brane system can be ensured. In addition, we take $ f(T)=T+\alpha T^2$ as an example to analyse the localization problem of the graviton zero mode. It is shown that the graviton zero mode can be localized on the brane.
\end{abstract}

%


\maketitle

\section{Introduction}

The extra-dimensional theory was first proposed to unify gravity and electromagnetic force \cite{Kaluza1921}. In the five-dimensional Kluza-Klein (KK) theory, the extra dimension is compacted into a circle with a Planck-scale radius and the electromagnetic field is generated from a part of a five-dimensional metric \cite{Kaluza1921,Klein1926}. In 1998, Arkani-Hamed, Dimopoulos, and Dvali (ADD) proposed a large extra dimension model in order to solve the hierarchy problem between the electroweak scale and the Planck scale \cite{ArkaniHamed:1998rs}. However, the ADD model introduces a new hierarchy between the size of the extra dimensions and the fundamental length. This problem can be solved by one of the subsequent Randall-Sundrum (RS) brane models \cite{Randall1999,Randall1999a}. The RS models have attracted much more attention since they were proposed, because of their great success in solving the hierarchy problem. However, the thickness of the RS branes is vanishing. The more realistic model is the so-called thick brane model, where the brane is constructed by the scalar field \cite{Goldberger:1999uk,Gremm2000a,DeWolfe2000a,Bazeia2009,Dzhunushaliev2010a}. Various of thick brane models have been investigated in Refs.~\cite{Charmousis:2001hg,Arias:2002ew,Barcelo:2003wq,Bazeia:2004dh,Navarro:2004di,CastilloFelisola:2004eg, Kanno:2004nr, BarbosaCendejas:2005kn,BarbosaCendejas:2007vp,Koerber:2008rx,deSouzaDutra:2008gm,Johnson:2008kc, Almeida:2009jc,Liu:2011wi,Chumbes:2011zt,Andrianov:2012ae,Kulaxizi:2014yxa,Zhong:2014kfk,Dutra:2014xla}.

General relativity only takes the effect caused by curvature. However, torsion can also lead
to the equivalent gravity theory known as the teleparallel equivalent of general relativity (TEGR) \cite{Aldrovandi2013}. The TEGR is constructed in the Weitzenb\"{o}ck manifold, which has vanishing curvature but non-vanishing torsion. The dynamical variables in the TEGR are the  vielbeins, which are defined in the tangent space at each point of the spacetime.

Since the difference between the torsion scalar $T$ and Ricci scalar $R$ is a boundary term $B$, the TEGR has the same field equations as general relativity. In order to find out what is the influence of torsion, one should modify the TEGR. Inspired by $f(R)$ gravity, the TEGR can be extended to $f(T)$ gravity, which was proposed to explain the acceleration of the universe by Bengochea and Ferraro \cite{Bengochea:2008gz}. Some models based on teleparallel gravity have been constructed in order to provide an alternative to inflation without inflation \cite{Ferraro:2006jd,Ferraro:2008ey}. Cosmological issues in higher-dimensional theories of $f(T)$ gravity have been discussed in Refs.~\cite{Bamba:2013fta,Fiorini:2013hva,Geng:2014}. The perturbation on cosmology in $f(T)$ gravity can be seen in Refs.~\cite{Chen:2010va,Izumi:2012qj, Haro:2013bea}.
In Ref. \cite{Bahamonde:2015zma}, the authors extended $f(T)$ gravity to more general $f(T,B)$ gravity and showed that $f(T)$ and $f(R)$ gravities are only its two special subcases.
Other related researches can be found for examples in Refs.~\cite{Bamba:2010wb,Sotiriou:2010mv,Ferraro:2011ks,Bamba:2011pz,Cai:2011tc,Bamba:2013rra, Wei:2015oua,Bamba:2015jqa}.

In this paper, we are interested in topics related to $f(T)$ brane scenario, i.e., brane model in $f(T)$ gravity.
Some solutions of $f(T)$ brane have been found recently in Refs. \cite{Yang:2012hu,Menezes:2014bta}.
The authors of Ref.~\cite{Yang:2012hu} considered the case of $f(T)=T+\alpha T^n$ and found that the brane will have a split with the increasing of the parameter $|\alpha|$, which is similar to the case of $f(R)$ brane \cite{Liu:2011wi,Bazeia:2013uva,Xu:2014jda}. In Ref. \cite{Menezes:2014bta}, $f(T)$ brane model with the same $f(T)=T+\alpha T^n$ was investigated with the first-order formalism for standard and generic Lagrangian densities of a scalar field. Analytical solutions with brane splitting were also found.
However, the stability of $f(T)$ brane system is still unknown. In this paper we will focus on this issue and analyse the localization of the graviton zero mode, which is an important issue in brane model since it will ensure the four-dimensional Newtonian potential on the brane~\cite{Kehagias:1999my,Garriga:1999yh,Csaki2000a,Bazeia:2003aw,Cvetic:2008gu,Zhong:2010ae, HerreraAguilar:2010kt,Cruz:2011kj,Santos:2012if,Ahmed:2012nh,Yang:2013a,Barbosa-Cendejas:2013cxa,Zhong:2014kfk, Fu:2014slg,Higuchi:2014bya,Gu:2014ssa}.

This paper is organised as follows. In Sec.~\ref{BraneworldModel}, we review the brane world model in $f(T)$ gravity briefly. In Sec.~\ref{TensorPerturbation}, we investigate the stability of the $f(T)-$brane system under tensor perturbation. The localization of the graviton zero mode is studied in Sec.~\ref{Localization}. The conclusion is given in the last section.

\section{Brane world model in $f(T)$ gravity}
\label{BraneworldModel}

We first give a brief review of teleparallel gravity. In teleparallel gravity, one usually works in the tangent space, which is associated with a spacetime point in the manifold, instead of the coordinate basis. At each point in the manifold with spacetime coordinates $x^M$, the vielbein fields $e_A(x^M)$ form an orthonormal basis of the tangent space. Capital Latin $A,B,C,\cdots=0,1,2,3,5$ and $M,N,O,\cdots=0,1,2,3,5$ label tangent space and spacetime coordinates, respectively. Clearly, $e_A (x^M)$ is a vector in the tangent space; and its components in a coordinate basis are ${e_A}^M$, which are both spacetime and Lorentz vectors.

The spacetime metric can be constructed from the  vielbein
\begin{equation}
g_{MN}={e^A}_M ~{e^B}_N \eta_{AB},
\label{Metricg}
\end{equation}
where $\eta_{AB}=\text{diag}(-1,1,1,1,1)$ is the Minkowski metric for the
tangent space. From the relation (\ref{Metricg}), we have
\begin{equation}
    {e_A}^{M}{e^A}_{N}=\delta^{M}_{N}, \qquad {e_A}^M {e^B}_M=\delta^{B}_{A}.
    \label{DeltaMN}
\end{equation}

It is the Weitzenb\"{o}ck connection $\tilde{\Gamma}^{\rho}_{~\mu\nu}$ rather than the  Levi-Civita connection $\Gamma^{\rho}_{~\mu\nu}$ that works in the teleparallel gravity, which is defined as
\begin{equation}
    \tilde{\Gamma}^{P}_{~MN}\equiv {e_A}^P\partial_N {e^A}_M.
\end{equation}
The asymmetric torsion tensor is constructed from the Weitzenb\"{o}ck connection:
\begin{equation}
   T ^{P}_{~MN}=\tilde{\Gamma}^{P}_{~NM}-\tilde{\Gamma}^{P}_{~MN}.
\end{equation}
The difference between the Weitzenb\"{o}ck connection and Levi-Civita connection is denoted by
\begin{equation} 
  K^{P}_{~MN}
  \equiv \tilde{\Gamma}^{P}_{~MN}-\Gamma^{P}_{~MN}
  =\frac{1}{2}\left(T^{~~P}_{M~N}+T_{N~M}^{~~P}-T^{P}_{~MN}\right).
  \label{KPMN}
\end{equation}
It is useful to define a tensor $S_{P}^{~MN}$:
\begin{equation} 
S_{P}^{~MN}\equiv\frac{1}{2}
   \left({K^{MN}}_{P}
   -{\delta^{N}_{P}{T^{QM}}_{Q}}
   +\delta^{M}_{P}{T^{QN}}_{Q}\right). \label{SPMN}
\end{equation}
Then the  Lagrangian of teleparallel gravity can be written as
\begin{eqnarray}
L_T=-\frac{M_*^3}{4}eT\equiv-\frac{M_*^3}{4}eS_{P}^{~~MN}T^{P}_{~~MN},
\label{Lagrangian}
\end{eqnarray}
where $e=\det({e^A}_M)=\sqrt{-g}$ with $g$ the determinant of the metric $g_{MN}$, and $M_*$ is fundamental Planck scale in five-dimensional spacetime. Similar to $f(R)$ gravity theory that generalizes general relativity, $f(T)$ gravity theory is the generalization of teleparallel gravity. Since the action of teleparallel gravity differs from that of general relativity only by a boundary term, it is equivalent to general relativity. However, $f(T)$ gravity is different from $f(R)$ gravity.

In this paper, we consider the brane world model in five-dimensional $f(T)$ gravity. The action is given by
\begin{equation}
  S=-\frac{M_*^3}{4}\int d^5x~ e~ f(T)+\int d^5x \mathcal{L}_M,
  \label{action}
\end{equation}
where $f(T)$ is a function of the torsion scalar $T$ and $\mathcal{L}_M$ stands for the matter Lagrangian density. We can get the filed equations by varying the action with respect to the  vielbein:
\begin{eqnarray}
&&e^{-1}f_T g_{NP}\partial_Q \left(e\,S_{\!M}^{~~PQ}\right)+f_{TT}S_{\!MN}^{~~~~Q}\partial_Q  T \nonumber\\ &&-{f}_{T}\tilde\Gamma^P_{~~QM}S_{\!PN}^{~~~~Q}+\frac{1}{4}g_{MN}f(T)=\mathcal{T}_{MN},
\label{field equation}
\end{eqnarray}
where $f_T\equiv\frac{\partial f(T)}{\partial T}$, $f_{TT}\equiv\frac{\partial^2 f(T)}{\partial T^2}$, and $\mathcal{T}_{MN}$ is the  energy-momentum tensor of the matter field. We have set $M_*=1$.
Note that, the above field equations only contain spacetime indices.

We would like to consider the flat brane world scenario, for which the metric ansatz reads as
\begin{equation}
   ds^2=e^{2A(y)}\eta_{\mu\nu}dx^\mu dx^\nu+dy^2,
   \label{metric}
\end{equation}
where $e^{2A(y)}$ is the warp factor and $\eta_{\mu\nu}=\text{diag}(-1,1,1,1)$ is the reduced metric on the brane world. Then the bulk  vielbein is ${e^A}_M=\text{diag}(e^{A},e^{A},e^{A},e^{A},1)$.
Here, the matter Lagrangian density is taken as $\mathcal{L}_M=e(-\frac{1}{2}\partial^M\phi~\partial_M\phi-V(\phi))$, where $\phi$ is a background scalar field that generates the brane. It is clear that the scalar field is a function of the extra dimension $y$, i.e., $\phi=\phi(y)$.
The explicit fields equations (\ref{field equation}) and the equation of motion of the scalar field are given by
\begin{eqnarray}
6A'^2 f_T+\frac{1}{4}f\!\!&=&\!\!
-V+\frac{1}{2}\phi'^2,  \label{EoMs1} \nn\\
 \frac{1}{4}f
+\left(\frac{3}{2}A{''}+6A'^2\right)f_T
-(36A'^2A{''})f_{TT}
\!\!&=&\!\!
-V-\frac{1}{2}\phi'^2,  \label{EoMs2}\nn\\
\phi{''}+4A'\phi'\!\!&=&\!\!\frac{dV}{d\phi},  \label{EoMsphi}
\end{eqnarray}
where prime denotes the derivation with respect to the extra dimension $y$.
Note that the above three equations are not independent.

\begin{figure*}[htb]
\begin{center}
\subfigure[$ $]  {\label{densityk}
\includegraphics[width=4.5cm]{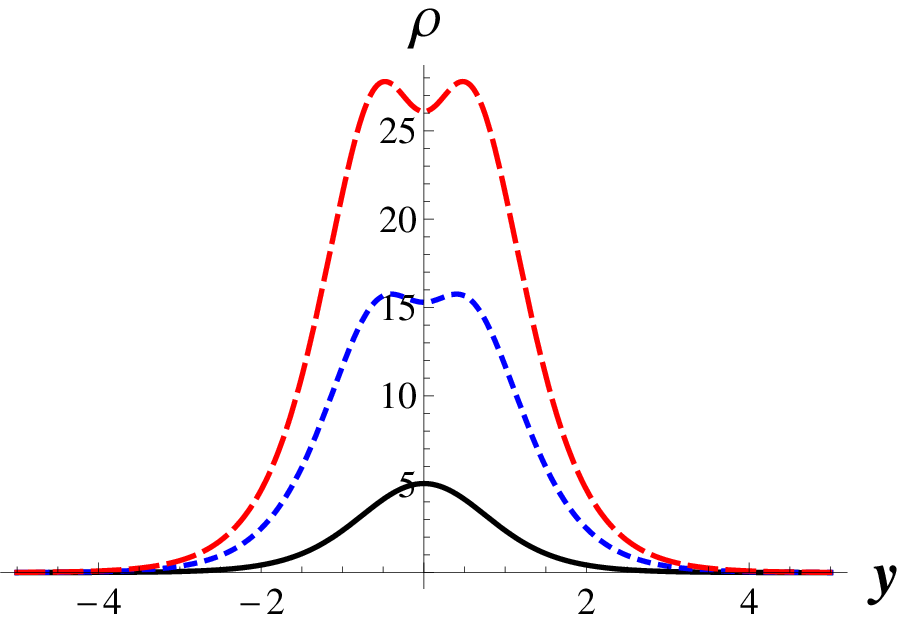}}
\subfigure[$ $]  {\label{vk}
\includegraphics[width=4.5cm]{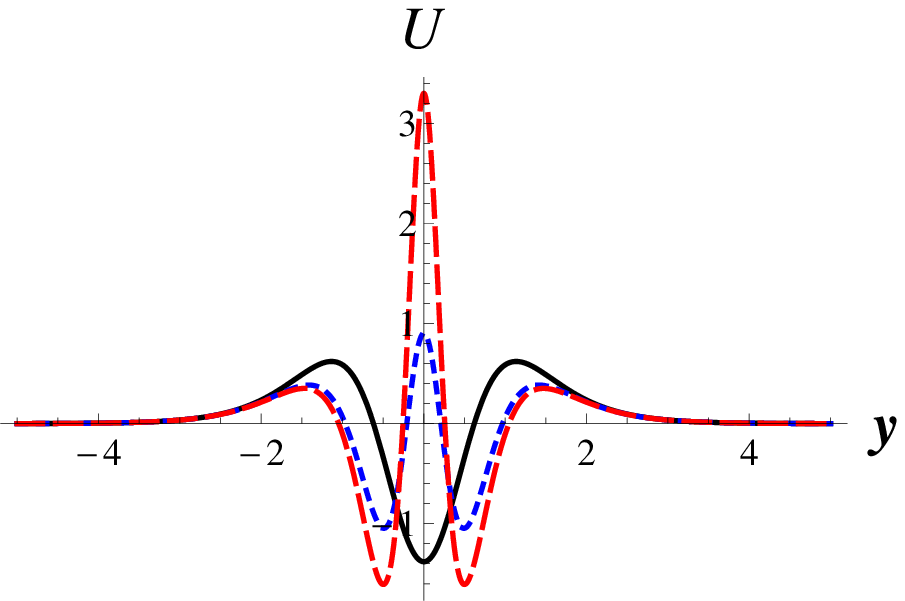}}
\subfigure[$ $]  {\label{zeromodek}
\includegraphics[width=4.5cm]{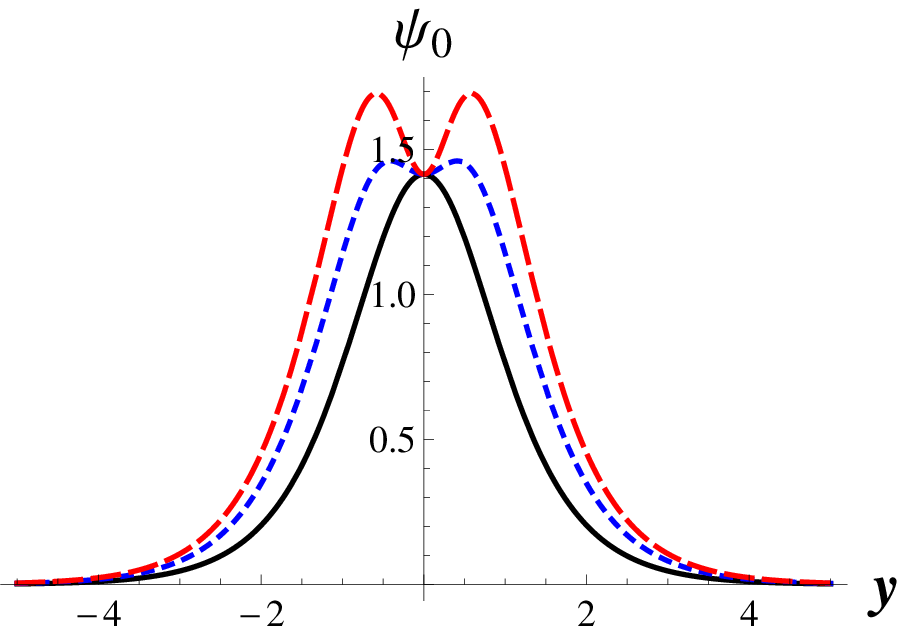}}
\end{center}
\caption{Plots of the energy density, effective potential, and zeromode for the brane solution of $f(T)=T+\alpha T^2$. The parameters are set to $b=1$, $\alpha=-0.005$ (the black solid lines),
$\alpha=-0.1$ (the blue dotted lines), and $\alpha=-0.2$ (the red dotted lines).}
\end{figure*}

\begin{figure*}[htb]
\begin{center}
\subfigure[$b=2$]  {\label{density1}
\includegraphics[width=4.5cm]{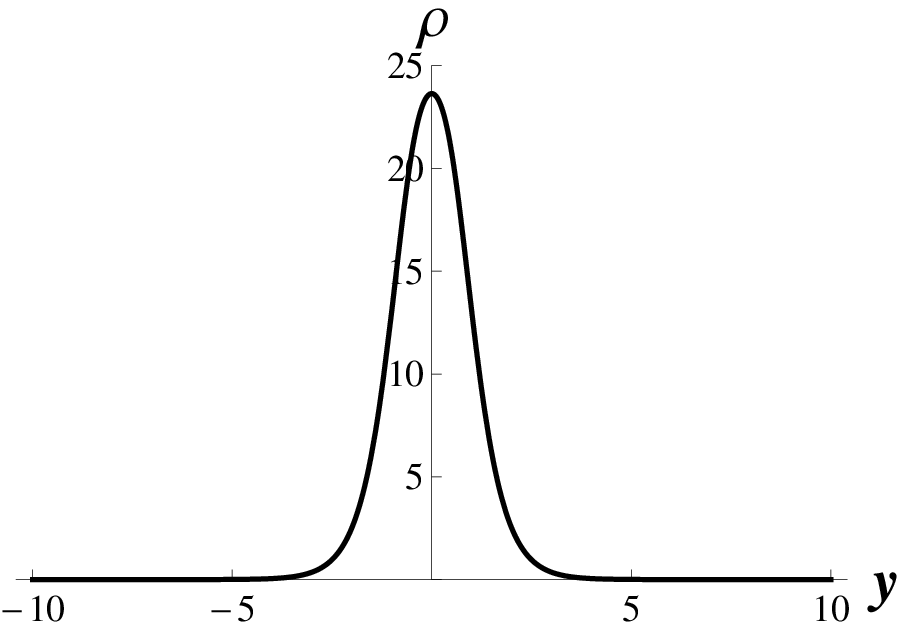}}
\subfigure[$b=20$]  {\label{density2}
\includegraphics[width=4.5cm]{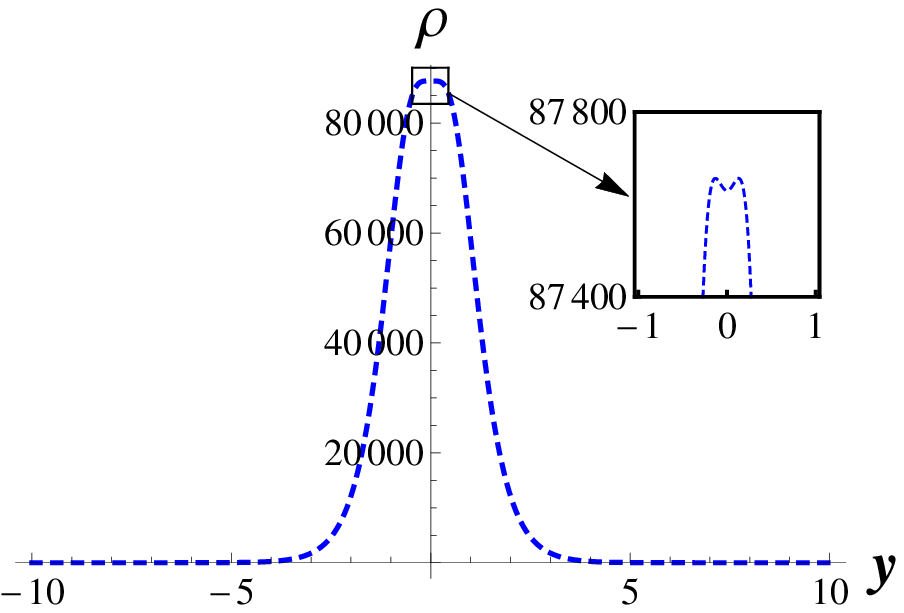}}
\subfigure[$b=40$]  {\label{density3}
\includegraphics[width=4.5cm]{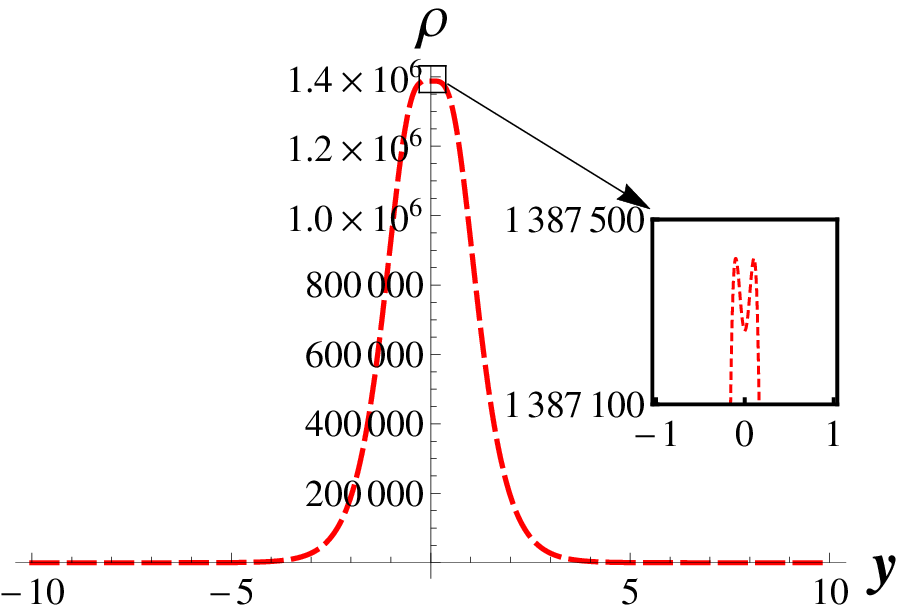}}
\end{center}
\caption{Plots of the energy density with $\alpha=-0.005$.}
~\label{density}
\end{figure*}

For $f(T)=T+\alpha T^n$, the thick brane solutions are investigated in Ref. \cite{Yang:2012hu}. Here, we list  two of them :
\begin{subequations}\label{s1}
\begin{eqnarray}
  e^{2A(y)}\!\!&=&\!\!\cosh^{-2b}(k y), \label{warpfactor1}\\
  \phi(y)\!\!&=&\!\!\sqrt{6b} \arctan\left(\tanh\Big(\frac{ky}{2}\Big)\right), \label{solution3}\\
  V(\phi)\!\!&=&\!\!\frac{3bk^2}{4}\left[(1+4b)\cos^2\left(\frac{2\phi}{\sqrt{6b}}\right)
                                  -4b
                             \right],
\end{eqnarray}
\end{subequations}
for $n=0$ or $n=1/2$, and
\begin{subequations}\label{s2}
\begin{eqnarray}
  e^{2A(y)}\!\!&=&\!\!\cosh^{-2b}(k y), \label{warpfactor2} \\
  \phi(y)\!\!&=&\!\!\sqrt{\frac{3b}{2}}
        \bigg(i\Big(\text{E}(i ky;u)
               -\text{F}(i ky;u)\Big) \nonumber\\
          &&+{\sqrt{\big(1+ u \sinh ^2(k y)\big)}}\tanh(ky)\bigg), \label{solution3}\\
  V(\phi(y))\!\!&=&\!\!
      \frac{3}{8} b k^2 \bigg(288 \alpha  b^3 k^2 \tanh ^4(k y)-8 b \tanh ^2(k y) \nonumber\\
          &&+ \text{sech}^4(k y) (u \cosh (2 k y)-u+2)\bigg),
    \label{solutionV}
\end{eqnarray}
\end{subequations}
for $n=2$, where $u=1-72 \alpha  b^2 k^2$.
Here $b$ and $k$ are positive parameters,
$F(y;q),~E(y;q)$ are the first and second kind elliptic integrals, respectively, and $\alpha \leq 1/(72k^2 b^2)$ for the second solution (\ref{s2}) in order to insure that the scalar field $\phi$ is real. When $y\rightarrow\pm\infty,~ A(y)\rightarrow -bk|y|$. Thus, the spacetime described by the above two solutions is asymptotically $AdS_5$. The cosmological constant is given by
\begin{eqnarray}
  \Lambda_5 \!\!\!&=&\!\!\! -3 b^2 k^2, ~~~~~~~~~~~~~~~~~~~~\text{for}~~~n=0,1/2. \\
  \Lambda_5 \!\!\!&=&\!\!\! -3 b^2 k^2 \left(1-36 \alpha  b^2 k^2\right) <0, ~~\text{for}~~~n=2.
\end{eqnarray}
The  energy densities for the above two brane solutions are respectively
\begin{eqnarray}
    \rho =\frac{3}{2}bk^2 \text{sech}(ky)+3b^2k^2 \text{sech}^2(ky), ~(n=0,1/2)
\end{eqnarray}
and
\begin{eqnarray}
    \rho\!\!\!&=&\!\!\!\Big(\frac{3}{2}bk^2
       +3b^2k^2\Big)\text{sech}^2(ky)
      -108\alpha b^4k^4\big[1-\tanh^4(ky)]\nonumber\\
      &&\!\!\!-108\alpha b^3k^4\tanh ^2(ky) \text{sech}^2(ky), ~(n=2)
\end{eqnarray}
from which we can see that each of the solutions stands for a thick brane localized near the origin of the extra dimension (see Fig. \ref{densityk} and \ref{density}), and the thickness of the brane is of about $1/k$.
We find that the brane will split when $1+2b+72\alpha<0$ and the split increases with $|1+2b+72\alpha|$.

In thick brane models, we usually take $M_*\sim k \sim M_{\text{Planck}}$, where $M_{\text{Planck}}$ is the effective four-dimensional Planck mass. In this paper, we set $M_*=k=1$.

\section{Tensor perturbation}\label{TensorPerturbation}

Next, we investigate the linear perturbations of the brane system. For simplicity, we only consider the transverse-traceless tensor perturbation in this paper, which is related to the gravitational wave and four-dimensional gravitons. The corresponding perturbed  vielbein reads as
\begin{eqnarray}
{e^A}_M=\left(
\begin{array}{cc}
e^{A(y)}({\delta^a}_\mu +{h^a}_\mu) & 0\\
0 & 1\\
\end{array}
\right).
\end{eqnarray}
According to the relation (\ref{DeltaMN}) and keeping only the linear order of the perturbation, we can get the inverse vielbein as
\begin{eqnarray}
{e_A}^M=\left(
\begin{array}{cc}
e^{-A(y)}({\delta_a}^\mu -{h_a}^\mu) & 0\\
0 & 1\\
\end{array}
\right),
\end{eqnarray}
where ${h_a}^\mu={h_a}^\mu(x^\mu,y)$.
Using Eq.~(\ref{Metricg}) we can easily get the corresponding metric:
\begin{eqnarray}
g_{MN}=\left(
\begin{array}{cc}
e^{2A(y)}(\eta_{\mu\nu}+\gamma_{\mu\nu}) & 0\\
0 & 1\\
\end{array}
\right),
\end{eqnarray}
and its inverse:
\begin{eqnarray}
g^{MN}=\left(
\begin{array}{cc}
e^{-2A}(\eta^{\mu\nu}-\gamma^{\mu\nu}) & 0\\
0 & 1\\
\end{array}
\right),
\end{eqnarray}
where
\begin{eqnarray}
\gamma_{\mu\nu}\!\!&=&\!\!({\delta^a}_\mu {h^b}_\nu+{\delta^b}_\nu {h^a}_\mu)\eta_{ab}, \nonumber \\
\gamma^{\mu\nu}\!\!&=&\!\!({\delta_a}^\mu {h_b}^\nu+{\delta_b}^\nu {h_a}^\mu)\eta^{ab},
\end{eqnarray}
and $\gamma^{\mu\nu}=\eta^{\mu\alpha}\eta^{\nu\beta}\gamma_{\alpha\beta}$,
${h_a}^\mu=\eta_{ab}\eta^{\mu\nu}{h^b}_\nu$.

Note that the TT tensor perturbation satisfies the following TT conditions:
\begin{equation}
\partial_\mu\gamma^{\mu\nu}=0=\eta^{\mu\nu}\gamma_{\mu\nu},
\end{equation}
whose equivalent  vielbein form is
\begin{eqnarray}
\partial_\mu({\delta_a}^\mu {h_b}^\nu+{\delta_b}^\nu{ h_a}^\mu)\eta^{ab}\!\!&=&\!\!0, \label{hTransverse}\\
{\delta_a}^\mu {h^a}_\mu\!\!&=&\!\!0. \label{hTraceless}
\end{eqnarray}
It is convenient to take the  vielbein form in the following work.

The nonvanishing components of the perturbed torsion tensors read as
\begin{eqnarray}
T^\rho_{~~5\mu}\!\!&=&\!\!A'\delta^\rho_\mu-A'({\delta^a}_\mu {h_a}^\rho-{\delta_a}^\rho {h^a}_\mu)+{\delta_a}^{\rho} {h'^a}_\mu,\nn\\
T^\rho_{~~\mu\nu}\!\!&=&\!\!{\delta_a}^\rho(\partial_\mu {h^a}_\nu-\partial_\nu {h^a}_\mu). \label{perturbedTorsionTensors}
\end{eqnarray}
In this paper, we always neglect the second-order terms for the perturbed quantities.
And the torsion scalar is
\begin{equation}
T=-12A'^2. 
\end{equation}
 Inserting the torsion tensor (\ref{perturbedTorsionTensors}) into Eqs.~(\ref{KPMN}) and (\ref{SPMN}), we get
\begin{eqnarray}
K^5_{~~\mu\nu}\!\!\!&=&\!\!e^{2A}\left(A'\eta_{\mu\nu}+A'\gamma_{\mu\nu}+\frac{1}{2}
\gamma'_{\mu\nu}\right),\nn \\
K^\rho_{~~5\nu}\!\!&=&\!\!-A'\delta^\rho_\nu-\frac{1}{2}{\delta^a}_\nu {{h'}_a}^\rho-\frac{1}{2}{\delta_a}^\rho {h'^a}_\nu,\nn \\
K^\rho_{~~\mu5}\!\!\!&=&\!\!\frac{1}{2}A'({\delta_a}^\rho {h^a}_\mu-{\delta^a}_\mu {h_a}^\rho)+\frac{1}{2}{\delta_a}^\rho {h'^a}_\mu-\frac{1}{2}{\delta^a}_\mu {{h'}_a}^\rho,\nn\\
K^\rho_{~~\mu\nu}\!\!\!&=&\!\!\!\frac{1}{2}{\delta^a}_\mu(\partial^\rho h_{a\nu}-\partial_\nu {h_a}^\rho)+\frac{1}{2}{\delta^a}_\nu(\partial^\rho h_{a\mu}-\partial_\mu {h_a}^\rho)\nn\\
&-&\!\!\frac{1}{2}{\delta_a}^\rho(\partial_\mu {h^a}_\nu-\partial_\nu {h^a}_\mu),
\end{eqnarray}
and
\begin{eqnarray}
S_5^{~~\mu\nu}\!\!&=&\!\!\frac{1}{2}e^{-2A}A'({\delta_a}^\mu h^{a\nu}-{\delta_a}^\nu h^{a\mu}) \nn\\
      &+&\!\!\frac{1}{4}e^{-2A}({\delta_a}^\mu h'^{a\nu}-{\delta_a}^\nu h'^{a\mu}),\nn  \\
S_\mu^{~~5\nu}\!\!&=&\!\!-\frac{3}{2}A'\delta^\nu_\mu+\frac{1}{4}{\delta_a}^\nu {h'^a}_\mu+\frac{1}{4}{\delta^a}_\mu {{h'}_a}^\nu,\nn\\
S_\mu^{~~\nu5}\!\!&=&\!\!\frac{3}{2}A'\delta_\mu^\nu-\frac{1}{4}{\delta_a}^\nu {h'^a}_\mu-\frac{1}{4}{\delta^a}_\mu {{h'}_a}^\nu,\nn\\
S_\mu^{~~\nu\rho}\!\!&=&\!\!\frac{1}{4}e^{-2A}{\delta_a}^\rho(\partial^\nu {h^a}_\mu-\partial_\mu h^{a\nu})+\frac{1}{4}{\delta^a}_\mu(\partial^\nu {h_a}^\rho-\partial^\rho {h_a}^\nu) \nn\\
\!\!&\!\!-\!\!&\!\!\frac{1}{4}e^{-2A}{\delta_a}^\nu(\partial^\rho {h^a}_\mu-\partial_\mu h^{a\rho})-\frac{1}{2}\delta_\mu^\nu e^{-2A}{\delta_a}^\theta\partial_\theta h^{a\rho}\nn\\
\!\!&\!\!+\!\!&\!\!\frac{1}{2}\delta^\rho_\mu e^{-2A}{\delta_a}^\theta\partial_\theta h^{a\nu},\nn\\
S_5^{~~5\nu}\!\!&=&\!\!-\frac{1}{2}e^{-2A}{\delta_a}^\rho\partial_\rho h^{a\nu},\nn\\
S_5^{~~\nu5}\!\!&=&\!\!\frac{1}{2}e^{-2A}{\delta_a}^\rho\partial_\rho h^{a\nu}.
\end{eqnarray}

The energy-momentum tensor of the scalar field is
\begin{equation}
\mathcal{T}_{MN}=\partial_M\phi\partial_N\phi-\frac{1}{2}g_{MN}g^{PQ}\partial_P\phi
\partial_Q\phi-g_{MN}V,
\end{equation}
where the perturbed scalar field can be expressed as
\begin{equation}
\phi=\bar\phi+\tilde\phi
\end{equation}
with $\bar\phi$ the background scalar field and $\tilde\phi=\tilde\phi(x^\mu,y)$ the perturbation.
Then we can get the perturbation of the energy-momentum tensor:
\begin{eqnarray}
\delta \mathcal{T}_{\mu\nu}\!\!&=&\!\!-e^{2A}
    \Big(\frac{1}{2}\bar{\phi}'^2\gamma_{\mu\nu}
         +\bar{\phi}'\tilde\phi'\eta_{\mu\nu} \nn\\
   && ~~~~~~+ V\gamma_{\mu\nu}
         + V_\phi\tilde\phi\eta_{\mu\nu} \Big),\nn\\
\delta \mathcal{T}_{5\mu}\!\!&=&\!\!\bar{\phi}'\partial_\mu\tilde\phi,\nn\\
\delta \mathcal{T}_{55}\!\!&=&\!\!\bar{\phi}'\tilde\phi'- V_\phi\tilde\phi.
\end{eqnarray}

Considering the traceless condition (\ref{hTraceless}), we have
\begin{equation}
\delta e=e e_A^M\delta e^A_M=e e^{-A}e^A\delta_a^\mu h^a_\mu=0.
\end{equation}
Then with  $\delta e=0$ and $\delta T=0$, the perturbed equations of (\ref{field equation})
are turned out to be
\begin{eqnarray}
&&e^{-1}f_T\delta g_{NP}\partial_Q\left(e S_M^{~~PQ}\right)+e^{-1}f_T g_{NP}\partial_Q
\left(e\delta S_M^{~~PQ}\right)\nn\\
&&+f_{TT}\delta S_{MN}^{~~~~Q}\partial_Q
T-f_T\delta\tilde\Gamma^Q_{~~PM}S_{QN}^{~~~~P}\nn\\
&&-f_T\tilde\Gamma^Q_{~~PM}\delta S_{QN}^{~~~~P}+\frac{1}{4}\delta g_{MN}f(T)=\delta \mathcal{T}_{MN}.
\end{eqnarray}
The $(\mu,5)$ components of the above perturbed equations vanish, and the $(\mu,\nu)$ components are given by
\begin{eqnarray}
&&f_T\Big( e^{-2A}\Box^{(4)}\gamma_{\mu\nu}
           -6A''\gamma_{\mu\nu}
           -24A'^2\gamma_{\mu\nu}
           +\gamma''_{\mu\nu}
      \Big)\nn\\
&&-24f_{TT}\big( A'A'' \gamma'_{\mu\nu}
                -6A'^2A'' \gamma_{\mu\nu}
           \big)
   - \gamma_{\mu\nu}f \nn\\
&&=\bar\phi'^2\gamma_{\mu\nu}+2
\bar\phi'\tilde\phi' \eta_{\mu\nu}
 +2 V \gamma_{\mu\nu}
 +2 V_{\phi} \tilde\phi \eta_{\mu\nu},~~
\label{pertubed field equation}
\end{eqnarray}
where $\Box^{(4)} \equiv \eta^{\mu\nu}\partial_{\mu}\partial_{\nu}$.
Besides, the $(\mu,\mu)$ component of the background field equations (\ref{field equation}) is
\begin{eqnarray}
144A'^2A''f_{TT}-6\left(A''+4A'^2\right)f_T-f
=\bar\phi'^2+2V. \label{mumucomponentbackgroundEOM}
\end{eqnarray}
By plugging Eq. (\ref{mumucomponentbackgroundEOM}) into Eq.~(\ref{pertubed field equation}), we obtain
\begin{eqnarray}
&&
\left(\gamma''_{\mu\nu}+4A'\gamma'_{\mu\nu}\right)f_T
  - 24A'A''\gamma'_{\mu\nu}f_{TT} \nn\\
&&=
 2 \eta_{\mu\nu}(\bar\phi'\tilde\phi'
   + V_{\phi}\tilde\phi).
\label{SimpliedPertubedFieldEquation}
\end{eqnarray}
The trace of Eq.~(\ref{SimpliedPertubedFieldEquation}) is
\begin{equation}
\bar\phi'\tilde\phi'+V_{\phi}\tilde\phi=0.
\end{equation}
Then, substituting the above equation to Eq.~(\ref{SimpliedPertubedFieldEquation}) yields the main
equation of the tensor perturbation:
\begin{eqnarray}
&& \left(e^{-2A}\Box^{(4)}\gamma_{\mu\nu}+\gamma''_{\mu\nu}+4A'\gamma'_{\mu\nu}\right)
   f_T \nn \\
 &&-24A'A''\gamma'_{\mu\nu}f_{TT}=0,
\label{mainequation}
\end{eqnarray}
which can also be rewritten as the following one if we introduce the relation $a=e^A$:
\begin{eqnarray}
 &&\left(a^{-2}\Box^{(4)}\gamma_{\mu\nu}+\gamma''_{\mu\nu}+4\frac{a'}{a} \gamma'_{\mu\nu}\right)f_T \nn\\
 && -24 \left(\frac{a'a''}{a^2}-\frac{a'^3}{a^3}\right)\gamma'_{\mu\nu}f_{TT}=0.
\label{perturbationEq1}
\end{eqnarray}
By making coordinate transformation
\begin{equation}
dz=e^{-A}dy,
\end{equation}
the tensor perturbation equation (\ref{perturbationEq1}) is transformed as
\begin{equation}
\left(\partial_z^2+2H\partial_z+\Box^{(4)}\right)\gamma_{\mu\nu}=0, \label{perturbationEq2}
\end{equation}
where
\begin{eqnarray}
H &=& \frac{3\partial_za}{2a}
     +12a^{-5}\Big(2(\partial_za)^3
     -a\partial_z^2a\partial_za\Big)\frac{f_{TT}}{f_T},\nn \\
  &=& \frac{3}{2}\partial_z A
       +12e^{-2A}\left(\left(\partial_z A \right)^3
                      -\partial_z^2 A\partial_z A
               \right)
        \frac{f_{TT}}{f_T}              .
\end{eqnarray}

Next, we introduce the KK decomposition
\begin{equation}
\gamma_{\mu\nu}(x^\rho,z)=\epsilon_{\mu\nu}(x^\rho) F(z) \psi(z), \label{KK decomposition}
\end{equation}
where $F(z)=e^{-\frac{3}{2}A(z)+\int{K(z)dz}}$ with
\begin{eqnarray}
K(z)&=&12a^{-5}\Big(a\partial_z^2 a\partial_z a-2(\partial_z a)^3\Big)\frac{f_{TT}}{f_T},\nn\\
    &=&12e^{-2A}\Big(\partial_z^2 A\partial_z A
                     -\left(\partial_z A \right)^3
               \Big)
        \frac{f_{TT}}{f_T}.
\end{eqnarray}
Then by substituting the KK decomposition (\ref{KK decomposition}) into the tensor perturbation equation
(\ref{perturbationEq2}), we get two equations: the KG-like equation for the four-dimensional KK
gravitons $\epsilon_{\mu\nu}$
\begin{eqnarray}
 \left(\Box^{(4)}+m^2\right)\epsilon_{\mu\nu}(x^\rho)= 0, \label{EoMepsilonMuNu}
\end{eqnarray}
and the Schr\"odinger-like equation for the extra-dimensional profile
\begin{eqnarray}
 \left(-\partial_z^2+U(z)\right)\psi = m^2\psi, \label{SchrodingerEquation}
\end{eqnarray}
where $m$ is the mass of the KK graviton and the effective potential is given by
\begin{eqnarray}
 U(z)=H^2+\partial_z H. \label{EffectivePotential}
\end{eqnarray}
The Schr\"odinger-like equation (\ref{SchrodingerEquation}) can be factorized as
\begin{equation}
\big(-\partial_z+H\big)\big(\partial_z+H\big)\psi=m^2\psi,
\end{equation}
which means that there is no four-dimensional graviton with $m^2<0$, and so any brane solution of $f(T)$
gravity theory is stable under the transverse-traceless tensor perturbation.
In order to localize a KK graviton on the brane, the corresponding extra-dimensional profile $\psi(z)$
(also called KK mode for simplify) should be satisfied the following normalization condition
\begin{equation}
 \int dz \;\psi^2(z) < \infty. \label{NormalizationCondition}
\end{equation}
The solution of the graviton zero mode (the four-dimensional massless graviton) is
\begin{equation}
\psi_0=N_0e^{\frac{3}{2}A-\int K(z)dz},\label{zeromodefunc}
\end{equation}
where $N_0$ is the normalization coefficient.

\section{Localization}\label{Localization}

In order to recover the four-dimensional gravity, the graviton zero mode should be localized on the brane.
In this section, we investigate the gravity localization problem by one of the brane solutions given in section \ref{BraneworldModel}.

We do not discuss the first brane solution with $n=\frac{1}{2}$, because the corresponding function
$f(T)=T+\alpha \sqrt{T}=-12A'^2+\alpha \sqrt{-12A'^2}$ is not real, so are the effective
 potential $U(z)$ (\ref{EffectivePotential}) and the graviton zero mode $\psi_0(z)$~(\ref{zeromodefunc}).

Now, we consider the second brane solution for $f(T)=T+\alpha {T}^2$, for which the effective potential in the physical coordinate $y$ is
\begin{eqnarray}
U(y)\!\!&=&\!\!
               \frac{1}{4} \text{sech}^{2 b}(y)
               \Bigg(\frac{32 b+16}{(1-24 b^2 \alpha) \cosh (2 y)+24 b^2 \alpha+1}\nn\\
\!\!& -&\!\!\frac{384 b^2 \alpha}{\big[\left(1-24 b^2 \alpha\right) \cosh (2 y)+24 b^2 \alpha+1\big]^2} \nn\\
\!\!&\!\!+\!\!&\!\!15 b^2-(3 b+2) (5 b+4) \text{sech}^2(y)\Bigg),
\label{potential}
\end{eqnarray}
and the graviton zero mode reads
\begin{equation}
\psi_0(y)=\frac{N_0 \sqrt{\left(1-24 b^2\alpha\right) \cosh (2 y)+24 b^2 \alpha+1}}{\cosh ^{\frac{3 b}{2}+1}(y)}.
\label{wavefunction}
\end{equation}
We can fix the normalization constant $N_0$ by the normalization condition
$\int^{+\infty}_{-\infty}\psi_0^2(z)dz=\int_{-\infty }^{+\infty } e^{-A(y)}\psi_0^2(y)dy=1$:
\begin{eqnarray}
N_0&=&\frac{2\left(24 b^2 \alpha-2 b-1\right)}{b (b+1)} \times \nn \\
&&\Big(1-\, _2F_1(1,-b-1;b+1;-1) \Big) ,
\end{eqnarray}
where $_2F_1$ is the Hypergeometric function. So, the graviton zero mode can be
 localized on the brane.

Now, we analyze the effect of the parameters $\alpha$ and $b$ on the localization of the graviton zero mode.
First, we consider $b=1$ and different values of $\alpha$.
As we can see from Fig.~\ref{vk}, when $\alpha$ tends to zero, the shape of the
effective potential is volcano like, and the graviton zero mode wave function has only one peak. This is an expected result that is consistent with the scenario of general relativity because the $\alpha T^2$ term can be neglected when $\alpha \rightarrow 0$.
When $|\alpha|$ increases, a new potential barrier located at $y=0$ and two potential wells around $y=0$ will appear. The height of the new potential barrier increases with the parameter $|\alpha|$. This is consistent with the change of the energy density with $|\alpha|$.
As a result, the  graviton zero mode wave function splits into two peaks as shown in
 Fig~\ref{zeromodek}, just like the split of the brane shown in Fig.~\ref{densityk}.

\begin{figure}[htb]
\begin{center}
\subfigure[$U(y)$]  {\label{vb}
\includegraphics[width=4cm]{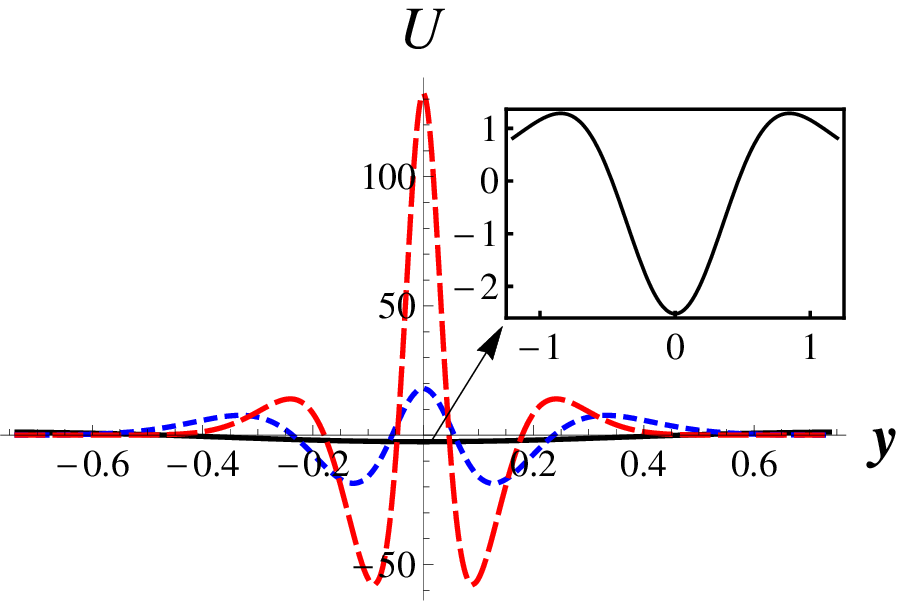}}
\subfigure[$\psi_0(y)$] {\label{zeromodeb}
\includegraphics[width=4cm]{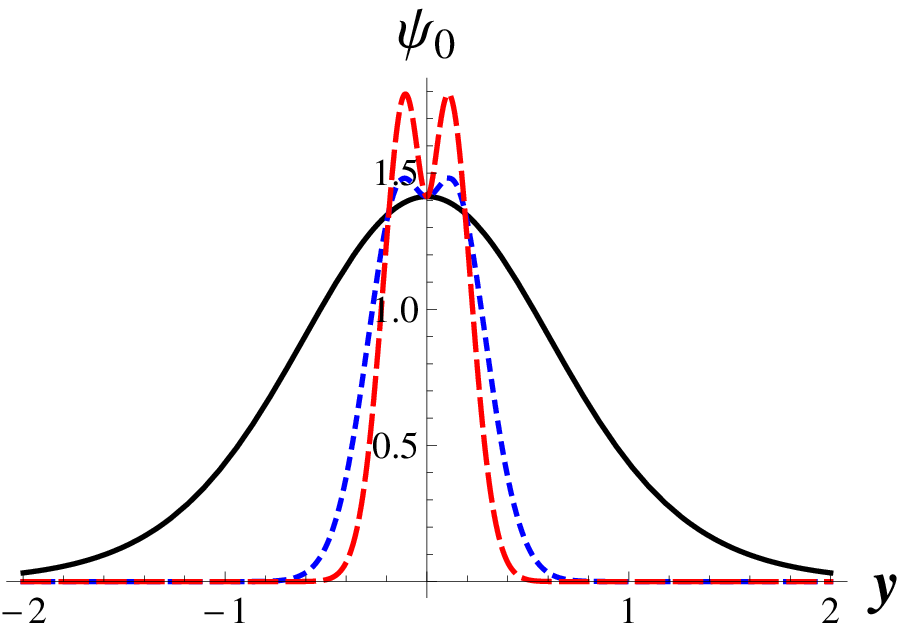}}
\end{center}
\caption{Plots of the effective potential and zero mode  for the brane solution of $f(T) =
T + \alpha T^2$. The parameters are set to $\alpha=-0.005$,
 $b=2$ (the black solid line), $b=20$ (the blue dotted line)
and $b=40$(the red dotted line).}
~\label{aaa}
\end{figure}

Next, we fix $\alpha=-0.005$ and take different values of $b$.
Figure \ref{aaa} shows the shapes of the effective potential and graviton
 zero mode.
 When $b$ increases, a barrier arises from
the volcano-like potential function (see Fig.~\ref{vb}), and the graviton zero mode splits into
 two peaks (see Fig.~\ref{zeromodeb}).
 The properties are similar to that of the case of fixed $b$.

\section{Conclusion}\label{conclusion}

In this paper, we investigated the tensor perturbation of the flat $f(T)$ thick brane model. By the relation between the perturbation of the metric and that of the vielbein we got the transverse-traceless condition in the form of the tensor perturbation of the  vielbein. Under that condition, we derived the tensor perturbation equation of the $f(T)$ brane. After the conformal transformation of the coordinate and the KK decomposition to the perturbation, we got a Schr\"odinger-like equation transformed from the main equation of the tensor perturbation equation.~(\ref{mainequation}), and proved that any solution for the $f(T)$ brane system is stable.

As an application, we studied one of the solutions for the $f(T)=T+\alpha T^2$ model, for which the brane has a little split. It was found that the graviton zero mode ( the four dimensional massless graviton) can be localized on the brane, which indicates that the four dimensional gravity can be recovered on the brane. Besides, by adjusting the parameters $\alpha$ and $b$, a new barrier appears in the effect potential for the KK modes, and correspondingly, the graviton zero mode also has a split. These results are related to the inner structure of the brane.

The scalar perturbations for $f(T)$ branes will leave for our future work.

%

\section{Acknowledgments}

This work was supported by the National Natural Science Foundation of China (Grants No. 11375075 and No. 11522541),
and the Fundamental Research Funds for the Central Universities (Grant No. lzujbky-2015-jl01).



\end{document}